\documentstyle[aps,prl,cite,epsfig,enumerate,verbatim]{revtex}
\draft
\begin{document}
\draft
\title{Geometric observation for the Bures fidelity
 between two states of a qubit}
\author{
 Jing-Ling Chen,$^{a,b}$ \footnote{Email: jinglingchen@eyou.com}
 Libin Fu,$^{a}$ Abraham A. Ungar,$^c$
 \footnote{Email: ungar@gyro.math.ndsu.nodak.edu}
 and Xian-Geng Zhao$^a$}
\address{
 $^a$
 Laboratory of Computational Physics,\\
 Institute of Applied Physics and Computational Mathematics, \\
 P.O. Box 8009(26), Beijing 100088, People's Republic of China. \\
 $^b$
 Department of Physics, Faculty of Science, National University of
 Singapore, \\ Lower Kent Ridge, Singapore 119260, Republic of
 Singapore. \\
 $^c$ Department of Mathematics, North Dakota State University,
 Fargo, North Dakota 58105, USA.}
 \maketitle

 \begin{abstract}

 In this Brief Report, we present a geometric observation for
 the Bures fidelity between two states of a qubit.

 \end{abstract}


 \section{Introduction}

 As is well known, the {\it trace distance} and the {\it Bures
 fidelity} are two important distance measures for quantum
 computation and quantum information [1-7]. A qubit is completely described
 by the $2\times 2$ density matrix as

 \begin{equation}
   \rho({\bf n}) = \frac{1}{2} ({\bf 1}+ {\vec \sigma} \cdot {\bf n}),
   \;\;\; |{\bf n}| \le 1.
  \label{t1}
 \end{equation}
 where ${\bf 1}$ is the unit matrix,
 ${\vec \sigma}=(\sigma_x, \sigma_y, \sigma_z)$ the Pauli
 matrices vector, and ${\bf n}$ the Bloch vector. $|{\bf n}|=1$
 corresponds to a pure state, otherwise a mixed state. Let

 \begin{eqnarray}
&&   \rho_1 = \frac{1}{2} ({\bf 1}+ {\vec \sigma} \cdot {\bf u}),\nonumber\\
&&   \rho_2 = \frac{1}{2} ({\bf 1}+ {\vec \sigma} \cdot {\bf v}),
 \label{t2}
 \end{eqnarray}
 be two states of a qubit. The trace distance and the Bures
 fidelity between $\rho_1$ and $\rho_2$ are given by the following equations
 \begin{eqnarray}
 &&   D(\rho_1,\rho_2)=\frac{1}{2} {\rm tr}|\rho_1 - \rho_2|,
 \label{t3}
 \end{eqnarray}
 \begin{equation}
 F(\rho_1,\rho_2) = \biggr[{\rm tr} \sqrt{ \sqrt{\rho_1} \rho_2 \sqrt{\rho_1}} \biggr]^2.
 \label{t4}
 \end{equation}
 One can easily obtain
  \begin{eqnarray}
 &&   D(\rho_1,\rho_2)=\frac{|{\bf u}-{\bf v}|}{2},
 \label{t5}
 \end{eqnarray}
 namely, the trace distance between two single qubit states has a
 simple geometric interpretation as half the ordinary Euclidean
 distance between points on the Bloch sphere. However, no
 similarly clear geometric interpretation is known for the Bures
 fidelity between two states of a qubit [7]. The purpose of this Brief
 Report is to provide a geometric observation for the Bures
 fidelity for the case of a qubit. In Sec. II, a definite
 geometric relation is formulated for the Bures fidelity in terms of
 {\it hyperbolic parameters}. Conclusion is made in the last section.

 \section{Formalism}

 {\bf Theorem:} The Bures fidelity between states $\rho_1$ and $\rho_2$ is equal to

 \begin{equation}
 F(\rho_1,\rho_2) = \frac{\cosh(\phi_{\bf w}/2)}
 {\cosh\phi_{\bf u}} \;
     \frac{\cosh(\phi_{\bf w}/2)}{\cosh\phi_{\bf v} },
 \label{t6}
 \end{equation}
 where $\phi_i$ $(i={\bf u}, {\bf v}, {\bf w})$ are rapidities.

  {\bf Proof:} Let us introduce the hyperbolic parameter $``\phi"$ to represent
 the Bloch vector as:
 \begin{equation}
  {\bf u}=\hat{\bf u} \tanh \phi_{\bf u},
   \label{t7}
 \end{equation}
 where $\hat{\bf u}={\bf u}/|{\bf u}|$ is a unit vector.
 It is easy to check $|{\bf u}| \le 1$ because of $|\tanh \phi_{\bf u}| \le
 1$; $\phi_{\bf u}=0$ corresponds to $|{\bf u}|=0$, while $\phi_{\bf u}
 \rightarrow \infty$ corresponds to $|{\bf u}|=1$. In other words, Eq. (\ref{t7})
 is a one-to-one mapping between $\phi_{\bf u}$ and ${\bf u}$.

 At this moment, the density matrix $\rho({\bf u})$ can be rewritten as
 \begin{equation}
  \rho({\bf u})= \frac{1}{2} ( {\bf 1}+
  {\vec \sigma} \cdot \hat{\bf u} \tanh \phi_{\bf u}).
   \label{t8}
 \end{equation}
 It is not difficult to observe that the relation between the density
 matrix $\rho({\bf u})$  and the Lorentz boost matrix
 $$L({\bf u})=\exp(\frac{\varphi_{\bf u}}{2} \vec{\sigma}\cdot\hat{\bf u})
 = {\bf 1} \cosh(\frac{\varphi_{\bf u}}{2})+ {\vec
 \sigma}\cdot{\hat{\bf u}}\sinh(\frac{\varphi_{\bf u}}{2})$$ is
 \begin{equation}
 \rho({\bf u})= \frac{L({\bf u})}{ 2 \cosh\phi_{\bf u}},\;\;\;
 \phi_{\bf u}  =\varphi_{\bf u}/2.
  \label{t9}
 \end{equation}
 Obviously, $\rho({\bf u})$ and $L({\bf u})$ are in one-to-one correspondence.
 For the former, the physical meaning of the vector ${\bf u}$ is the Bloch
 vector in quantum mechanics, while for the latter the relativistic velocity.
 Due to the rapidity $\varphi$, i.e., the hyperbolic angle, special relativity
 can be formulated in terms of hyperbolic geometry. Consequently, some physical
 quantities have been found to have geometric meanings, such as the
 Thomas rotation angle (sometimes also called the Wigner angle) corresponds to the
 defect of a hyperbolic triangle [8,9]. Since
 $\rho({\bf u})$ and $L({\bf u})$ are in one-to-one
 correspondence, we are led to view the Bloch vector ${\bf u}$ as an analogous
 relativistic velocity, and the angle $\phi$ as the rapidity. Motivated by this, we
 will find out a geometric interpretation for the quantum fidelity $F(\rho_1,\rho_2)$
 in the framework of hyperbolic geometry.

  From the Addition Law of Velocities in special relativity:
  \begin{equation}
   {\bf w}= {\bf u} \; {\oplus} \; {\bf v}  =
   \frac{1}{1+ \frac{ {\bf u} \cdot {\bf v} } {c^2} }
    \biggr[ {\bf u} + \frac{1}{\gamma_{\bf u}}{\bf v} +
    \frac{1}{c^2} \frac{\gamma_{\bf u}}{1+\gamma_{\bf u}}
    ({\bf u} \cdot {\bf v}) {\bf u}
    \biggr],
 \label{t10}
 \end{equation}
 where $\gamma_{\bf u}=1/\sqrt{1-|{\bf u}|^2/c^2}$ is the Lorentz factor
 and $c$ is the speed of light in vacuum space, we have
 \begin{equation}
   \gamma_{\bf w}= \gamma_{\bf u} \gamma_{\bf v}(1+ {\bf u} \cdot {\bf v}),
 \label{t11}
 \end{equation}
 or
 \begin{equation}
   \cosh\phi_{\bf w}= \cosh\phi_{\bf u} \cosh\phi_{\bf v}
   (1+ \hat{\bf u} \cdot \hat{\bf v} \tanh\phi_{\bf u} \tanh\phi_{\bf v}),
 \label{t12}
 \end{equation}
 which is the Cosin-Law in the hyperbolic geometry.

 From Eq.(\ref{t9}), one obtains
 \begin{equation}
 \sqrt{\rho({\bf u})}= \frac{\cosh(\phi_{\bf u}/2)}
 {\sqrt{2 \cosh \phi_{\bf u}}}
 [{\bf 1} + {\vec \sigma} \cdot \hat{\bf u} \tanh(\phi_{\bf u}/2)].
  \label{t13}
 \end{equation}
 From $ {\rm det}(\sqrt{\rho_1} \rho_2 \sqrt{\rho_1}-\Lambda {\bf 1})=0$,
 we have
 \begin{eqnarray}
 && \Lambda^2 - \frac{\gamma_{\bf w}}{2 \gamma_{\bf u} \gamma_{\bf v}}
  \Lambda + \frac{1}{16 \gamma_{\bf u}^2 \gamma_{\bf v}^2}=0,
  \label{t14}
 \end{eqnarray}
 so that
 \begin{eqnarray}
 && \Lambda_\pm= \frac{ \cosh\phi_{\bf w} \pm \sinh\phi_{\bf w} }
 {4 \cosh\phi_{\bf u} \cosh\phi_{\bf v}}.
  \label{t15}
 \end{eqnarray}
 Thus the Bures fidelity is
 \begin{equation}
 F(\rho_1,\rho_2)= (\sqrt{\Lambda_+}+ \sqrt{\Lambda_-})^2=
 \frac{\cosh(\phi_{\bf w}/2)}
 {\cosh\phi_{\bf u}} \;
     \frac{\cosh(\phi_{\bf w}/2)}{\cosh\phi_{\bf v} }.
  \label{t16}
 \end{equation}
 This ends the proof.

 \section{Conclusion}

    In Fig. 1, we draw a hyperbolic triangle $\Delta ABC$
 formed by three hyperbolic angles
 $\{ \phi_{\bf u}=|AB|,\phi_{\bf v}=|AC|,\phi_{\bf w}=|BC| \}$,
 where $D$ is the midpoint of the side $BC$. As one can see that
 the trace distance $D(\rho_1, \rho_2)=|{\bf u}-{\bf v}|/2$ is
 related to an ordinary Euclidean triangle, whose three sides are
 $|{\bf u}|$, $|{\bf v}|$ and $|{\bf u}-{\bf v}|$; similarly,
 the Bures fidelity is related to a hyperbolic triangle, it is a
 multiplication of the ratio $\cosh(\phi_{\bf w}/2)/\cosh\phi_{\bf u}$
 and the ratio $\cosh(\phi_{\bf w}/2)/\cosh\phi_{\bf v}$. From
 Eq. (\ref{t16}), one easily sees that $F(\rho_1, \rho_2)$ is
 symmetric in its inputs, i.e., $F(\rho_1, \rho_2)=F(\rho_2,
 \rho_1)$, and is invariant under unitary transformations on the
 state space.

 \begin{figure}
 \centerline{\epsfig{file=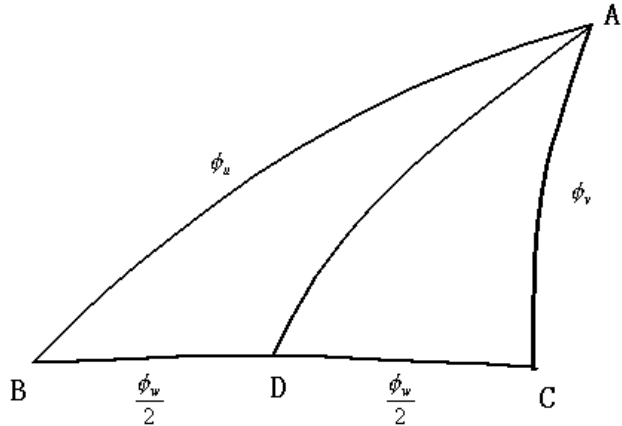,width =.5\linewidth}}
 \caption{The hyperbolic triangle $\Delta ABC$. Its three sides
 are
 $|AB|=\phi_{\bf u}=\tanh^{-1}|{\bf u}|$,
 $|AC|=\phi_{\bf v}=\tanh^{-1}|{\bf v}|$,
 $|BC|=\phi_{\bf w}=\tanh^{-1}|{\bf w}|$.
 $D$ is the midpoint of the side $BC$. The angle between AB and AC
 is equal to $\pi-\arccos(\hat{\bf u} \cdot \hat{\bf v})$.}
 \end{figure}

     In conclusion, we have presented a geometric observation
 for the Bures fidelity between two states of a qubit. It is also
 interesting and significant to study the geometric meaning of
 the Bures fidelity for the case of a qu$N$it (i.e., an $N$-dimensional
 quantum object, $N=2$ corresponds to a qubit) [10], since the
 calculation becomes much more complicated, we shall investigate it
 elsewhere. Nevertheless, we believe that a similar simple hyperbolic
 geometric relation, such as Eq.(\ref{t16}), is possibly held for the
 case of a qu$N$it.


\begin{thebibliography}{99}

\bibitem{1}  D. Bures, Trans. Am. Math. Soc. {\bf 135}, 199 (1969);
             A. Uhlmann, Rep. Math. Phys. {\bf 9}, 273 (1976);
             Rep. Math. Phys. {\bf 24}, 229 (1986);
             Ann. Phys. Leipzig {\bf 46}, 63 (1989);
             M. Hubner, Phys. Lett. A {\bf 163}, 229 (1992);
             {\bf 179}, 226 (1993).
\bibitem{2}  R. Jozsa, J. Mod. Opt. {\bf 41}, 2315 (1994);
             B. Schumacher, Phys. Rev. A {\bf 51}, 2738 (1995).
\bibitem{3}  M.B. Ruskai, Rev. Math. Phys. {\bf 6}, 1147 (1994);
             H. Barnum, C.M. Caves, C.A. Fuchs, R. Jozsa, and
             B. Schumacher, Phys. Rev. Lett. {\bf 76}, 2828 (1996).
             C.A. Fuchs, Ph.D. thesis, The University of New Mexico,
             Albuquerque, NM, 1996. arXive e-print quant-ph/9601020.
\bibitem{4}  V. Vedral, M.B. Plenio, M.A. Rippin, and P.L. Knight,
             Phys. Rev. Lett. {\bf 78}, 2275 (1997).
\bibitem{5}  J. Twamley, J. Phys. A {\bf 29}, 3723 (1996);
             H. Scutaru, J. Phys. A {\bf 31}, 3659 (1998);
             X.B. Wang, C.H. Oh, and L.C. Kwek, Phys. Rev. A {\bf
             58}, 4186 (1998); L.C. Kwek, C.H. Oh, X.B. Wang, and
             Y. Yeo, Phys. Rev. A {\bf 62}, 052313 (2000).
\bibitem{6}  B.W. Schumacher, Phys. Rev. A {\bf 54}, 2614 (1996);
             E. Knill and R. Laflamme, Phys. Rev. A {\bf 55}, 900
             (1997); H. Barnum, E. Knill, and M.A. Nielsen,
             arXive e-print quant-ph/9809010.
\bibitem{7}  M.A. Nielsen and I.L. Chuang, {\it Quantum
             Computation and Quantum Information}, Cambridge University
             Press, 2000, pp. 399-424. http://www.cambridge.org
\bibitem{8}  J.-L. Chen and M.-L. Ge, J. Geom. Phys.
             {\bf 25}, 341 (1998); P.K. Aravind, Am. J. Phys.
             {\bf 65}, 634 (1997);
             A.A. Ungar, Found. Phys. {\bf 27}, 881 (1997);
             J.-L. Chen and A.A. Ungar, Found. Phys.
             {\bf 31}, 1611 (2001).
\bibitem{9}  Abraham A. Ungar, {\it Beyond the Einstein addition law and
             its gyroscopic Thomas precession: the theory of gyrogroups
             and gyrovector spaces}, (Kluwer Academic
             Publishers, Dordrecht, 2001), pp. 253-278.
\bibitem{10} D. Kaszlikowski, P. Gnaci\'nski, M. \.Zukowski,
             W. Miklaszewski and
             A. Zeilinger, Phys. Rev. Lett. {\bf 85}, 4418 (2000);
             J.-L. Chen, D. Kaszlikowski, L.C. Kwek, C.H. Oh and
             M. \.Zukowski, Phys. Rev. A {\bf 64}, 052109 (2001).

\end{thebibliography}
 \end{document}